\documentclass{PoS}

\title{The nature of LS 5039 under the scrutiny of gamma-rays}

\ShortTitle{The nature of LS 5039}

\author{\speaker{Diego F. Torres}\\
        ICREA \& Institut de Ciencies de l'Espai (IEEC-CSIC), Campus UAB, Fac. de Ciencies, Torre C5, parell, 2a planta, 08193
  Barcelona\\
        E-mail: \email{dtorres@ieec.uab.es}}

\author{Agnieszka Sierpowska-Bartosik\\
        Institut de Ciencies de l'Espai (IEEC-CSIC),
  Campus UAB, Fac. de Ciencies, Torre C5, parell, 2a planta, 08193
  Barcelona,  Spain. \\
        E-mail: \email{agni@ieec.uab.es}}

\abstract{Several gamma-ray binaries have been recently detected by the High-Energy Stereoscopy Array (H.E.S.S.) and the Major Atmospheric Imaging Cerenkov (MAGIC) telescope. In two cases, their nature is unknown, since a final observational feature for a black hole or a pulsar compact object companion is still missing. One such system is LS 5039. Here we present results from a model (it includes a detailed account of the system geometry, the angular dependence of processes such as Klein-Nishina inverse Compton and gamma-gamma absorption, and a Monte Carlo simulation of cascading) of the high energy phenomenology of LS 5039 in which it is assumed that the companion object is a pulsar rotating around an O6.5V star in the 3.9 days orbit. We show that the H.E.S.S. phenomenology at all scales (spectra along the orbit in both broad and short phase-bins and lightcurve) is described within this model. We focus on presenting predictions for photons with lower energies (for E>1  GeV), subject to test in the forthcoming months with the GLAST satellite, and we also present predictions for future observations with high-energy arrays, such as H.E.S.S. II. Both set of predictions go beyond the description of the current data, and could
provide a high-energy determination of the system's composition.
}

\FullConference{VII Microquasar Workshop: Microquasars and Beyond\\
		 September 1-5 2008\\
		 Foca, Izmir, Turkey}

\begin{document}

\section{Prologue}

Very recently, a few massive binaries have been identified as variable very-high-energy (VHE) $\gamma$-ray sources. They are
PSR B1259-63 (Aharonian et al. 2005a), LS 5039 (Aharonian et al. 2005b, 2006), LS I +61 303 
(Albert et al. 2006, 2008a,b), and Cyg X-1 (Albert et al. 2007). 
The nature of only two of these binaries is considered known: PSR B 1259-63 is formed with a pulsar whereas Cyg X-1 is formed with a black hole. The nature of the two remaining systems is under discussion. The high-energy phenomenology of 
Cyg X-1 is different from that of the others. It has been detected just once in a flare state for which a duty cycle is yet unknown and its SED does not qualifies it as a gamma-ray binary. The three other sources, instead, present a behavior that is fully correlated with the orbital period and the gamma-ray band dominates their SEDs.
In recent papers (Sierpowska-Bartosik and Torres 2007, 2008a,b) under the assumption that LS 5039 is composed by a pulsar rotating around an O6.5V star in the $\sim 3.9$ day orbit,  we have
presented  a leptonic theoretical modeling (it includes a detailed account of the system geometry, the angular dependence of processes such as Klein-Nishina inverse Compton and gamma-gamma absorption, and a Monte Carlo simulation of cascading) for the high-energy phenomenology observed by  H.E.S.S. These works studied the lightcurve and the spectral orbital variability in both broad orbital phases and in shorter (0.1 phase binning) timescales. We have also analyzed how this model could be tested by Gamma-ray Large Area Space Telescope (GLAST), and future TeV observations. Details of the model and its implementation concerning the binary geometry, wind termination,  opacities to different processes along the orbit of the system are fully given in these previous works. Here, we summarize some of our results.

\section{Geometry and opacities}

\begin{figure*}
  \centering
    \includegraphics*[width=0.5\textwidth,angle=0,clip]{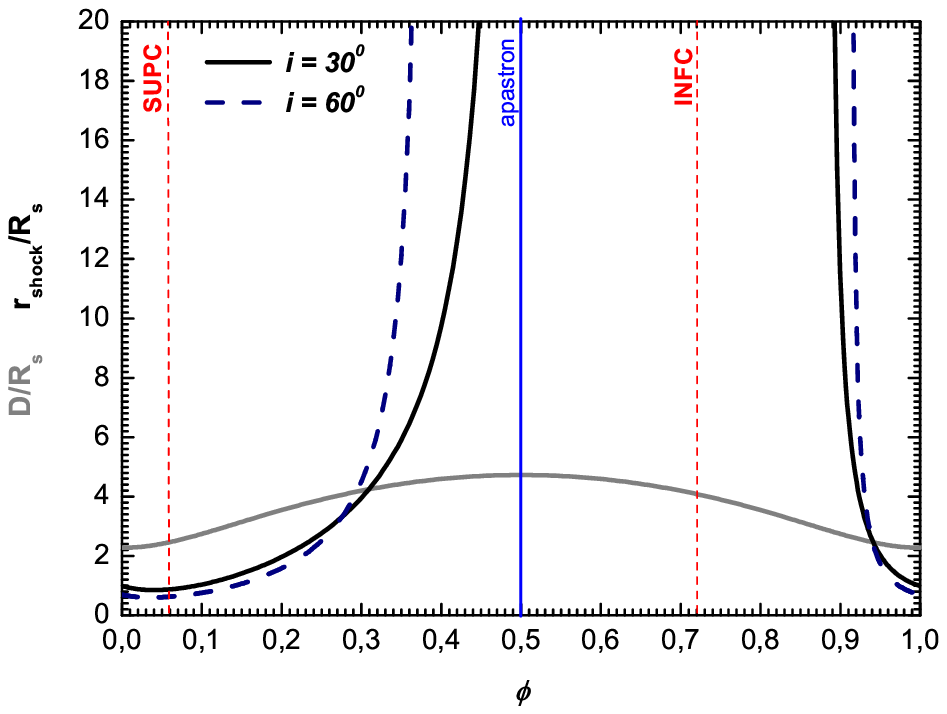}
   \includegraphics[width=0.49\textwidth,angle=0,clip]{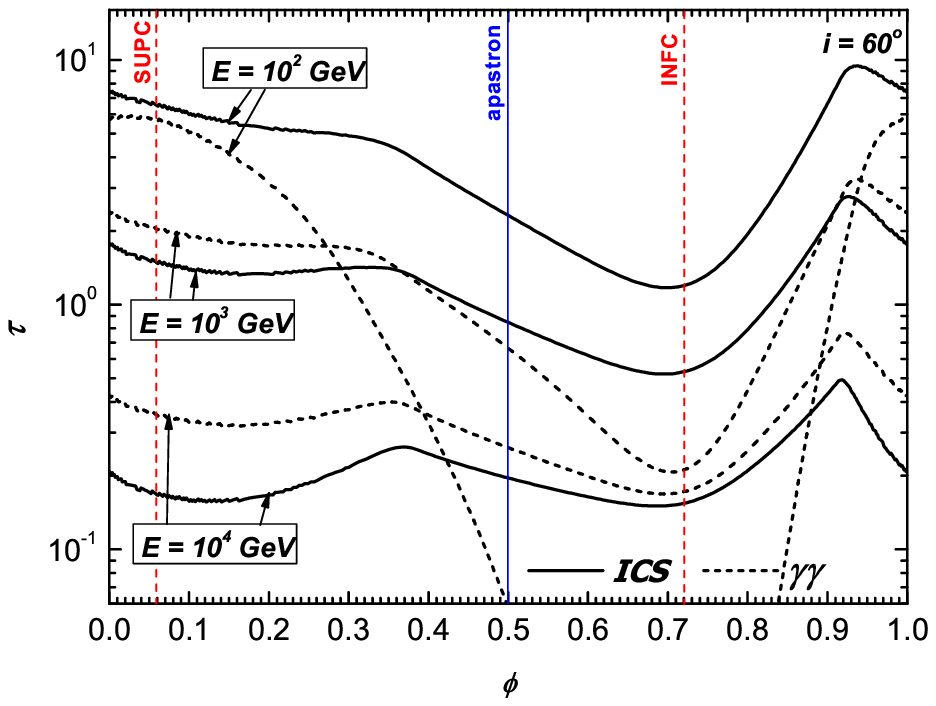}
\caption{\label{fig:tau_phi} {Left: The distance from the pulsar  to the termination shock in the direction to the observer (in units of stellar radius $R_s$), for the two different inclination angles, $\textit{i}$, analyzed in this paper. INFC, SUPC, periastron, and apastron phases are marked. Additionally, the gray line shows the separation of the binary (also in units of $R_s$) as a function of phase along the orbit.
Right: Opacities (in the direction to the observer) to ICS for electrons and pair production for photons as a function of the orbital phase, $\phi$ in the LS 5039 system (for an example with an inclination of the binary orbit of $i=60^0$) and energy of injected particle. }}
\label{go}
\end{figure*}

In the context of LS 5039 properties,  the dependence of the distance from the pulsar to the termination shock in the direction to the observer varies much as a function of the orbital phase. This is shown in Fig. \ref{go}. 
We find that for both inclinations considered, the wind is unterminated for a specific range of phases along the orbit, i.e., the electron propagating in the direction of the observer would find no shock.
The PWZ would always be limited in the observer's direction only if the inclination of the system is close to zero, i.e.,  the smaller the binary inclination the narrower the region of the wind non-termination viewed by the observer. 
The unterminated wind is viewed by the observer at the phase range between apastron and INFC, while a strongly limited wind appears from periastron to SUPC. 
Note that the important differences in the range of phases in which the unterminated wind appears for distinct inclinations (e.g., the observer begins to see the unterminated pulsar wind at $\sim 0.36$ for $i=60^o$ compared to  $\sim 0.45$ for  $i=30^0$) may produce a distinguishable feature of any model with a fixed orbital inclination angle. 

The conditions for leptonic processes for this specific binary can be discussed based on optical depths to ICS and $\gamma\gamma$ absorption. The target photons for IC scattering of injected electrons and for $e^+e^-$ pair production for secondary photons are low energy photons of black-body spectrum with temperature $T_s = 3.9 \times 10^4$, which is the surface temperature of the massive star. This is an anisotropic field as the source of thermal photons differ from the place of injection of relativistic electrons, which for simplicity is assumed to be at the pulsar location for any given orbital phase.
For fixed geometry parameters, the optical depths change with the separation of the binary (in general with the distance to the massive star), the angle of injection (the direction of propagation with respect to the massive star), and the energy of injected particle (the electrons or photons for $\gamma$ absorption).
The dependence of the optical depths upon the orbital phase for specific parameters of the LS 5039 is shown in Fig. \ref{go}. These opacities are calculated up to the termination shock in the observer direction (see Sierpowska-Bartosik and Torres 2008a for a more generic computation of opacities, up to infinity). The presence of the shock (at a distance $r_s$) limits the optical depth for IC. Since this parameter is highly variable along the orbit, its influence on the optical depth values is not minor. 
As the angle to the observer, which defines the primary injection angle, vary in the range $(90-i, 90+i)$ (for INFC and SUPC, respectively), we found that there is a range of orbital phases for which the PWZ is non-terminated. In that case the cascading process --which develops linearly-- is followed up to electron's complete cooling (defined by the energy $E_{min} = 0.5\, \rm GeV$).  
Apart from the condition for full electron cooling, in the case of the terminated wind the cascade is followed up to the moment when the electron reaches the shock region, whatever happens first. In all cases, photon absorption is computed thereafter.

\section{Predictions and comparison with observations}

\begin{figure*}[t]
\centering
\includegraphics[width=.49\textwidth]{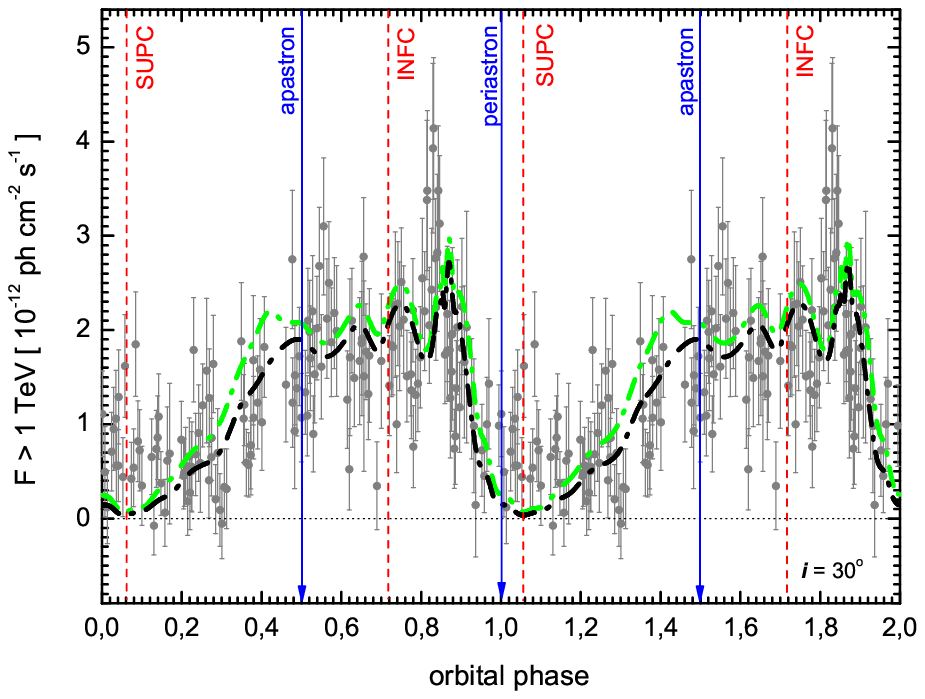}
\includegraphics[width=.49\textwidth]{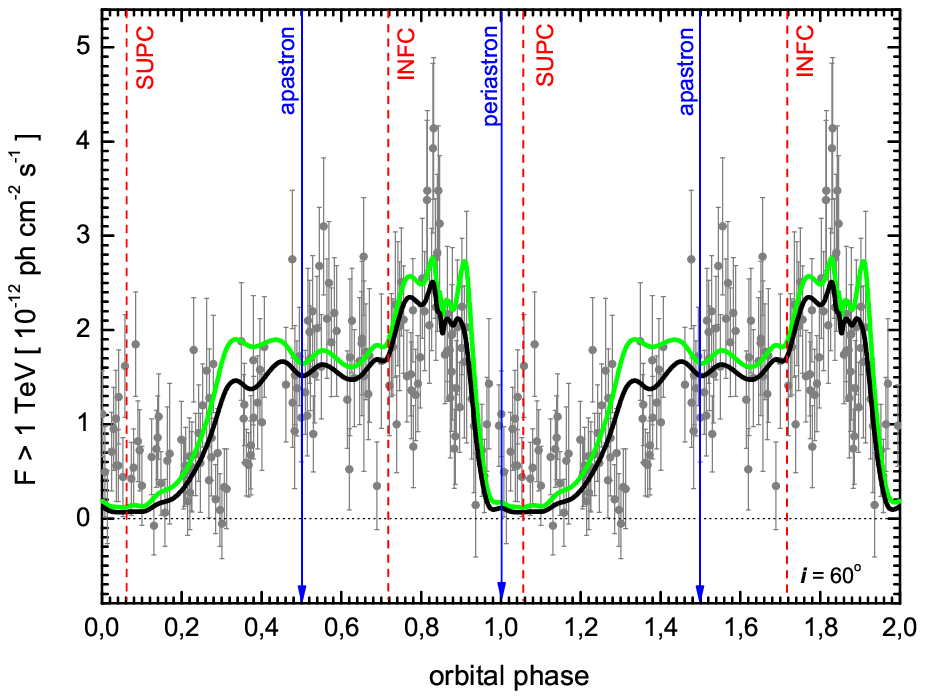}\\
\includegraphics[width=.49\textwidth,trim=0 5 0 10]{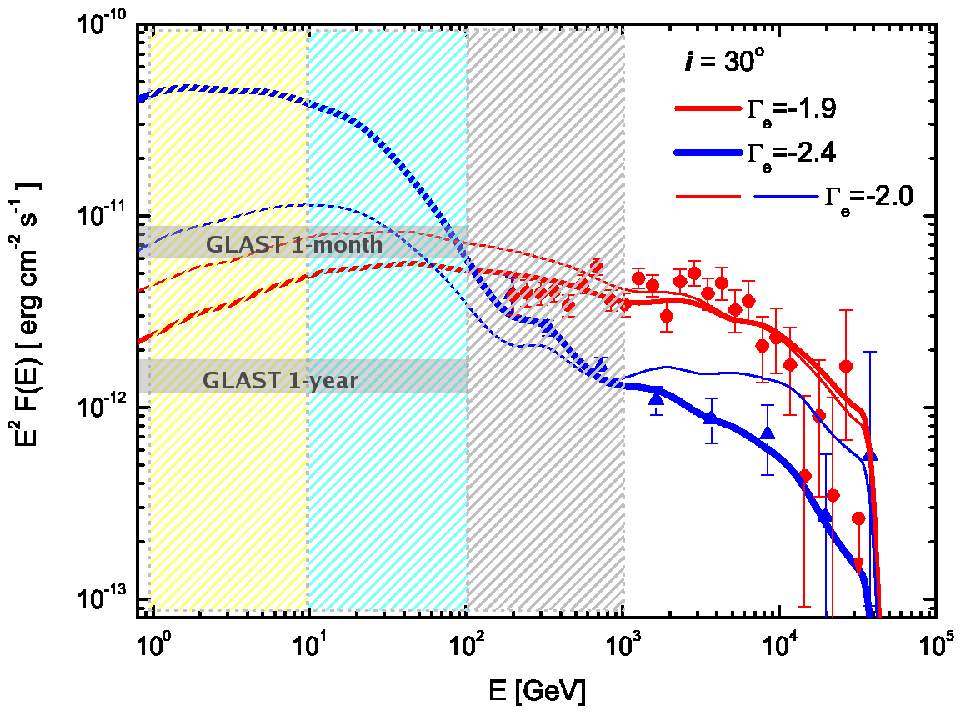}
\includegraphics[width=.49\textwidth, trim=0 5 0 10]{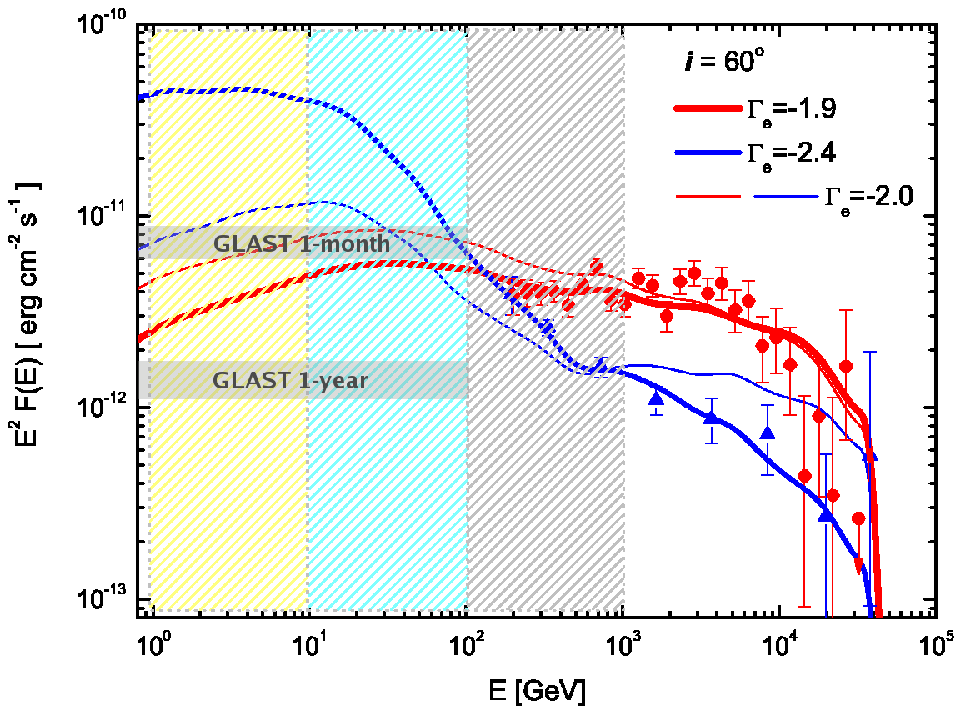}\\
\includegraphics[width=.49\textwidth]{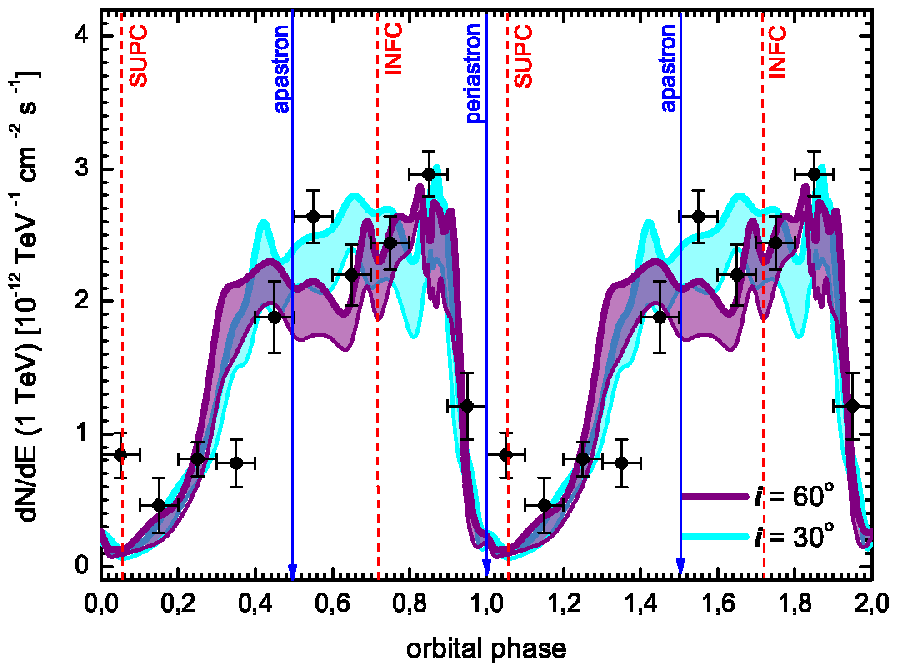}
\includegraphics[width=.49\textwidth]{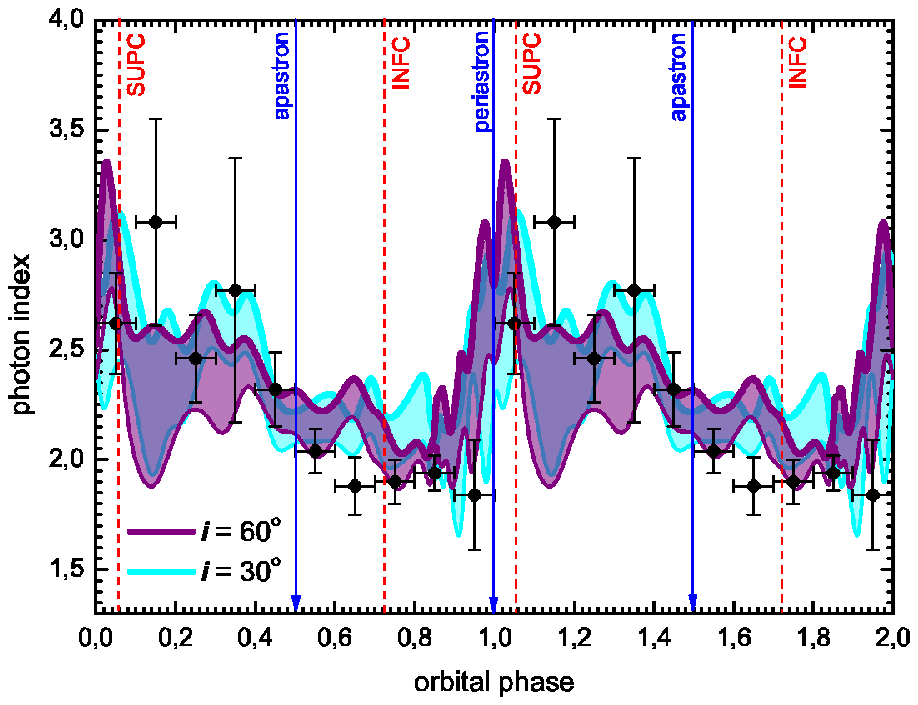}
\caption{Results of the model, see text for an explanation}
\label{LC}
\end{figure*}

\begin{figure*}
\includegraphics[width=.5\textwidth]{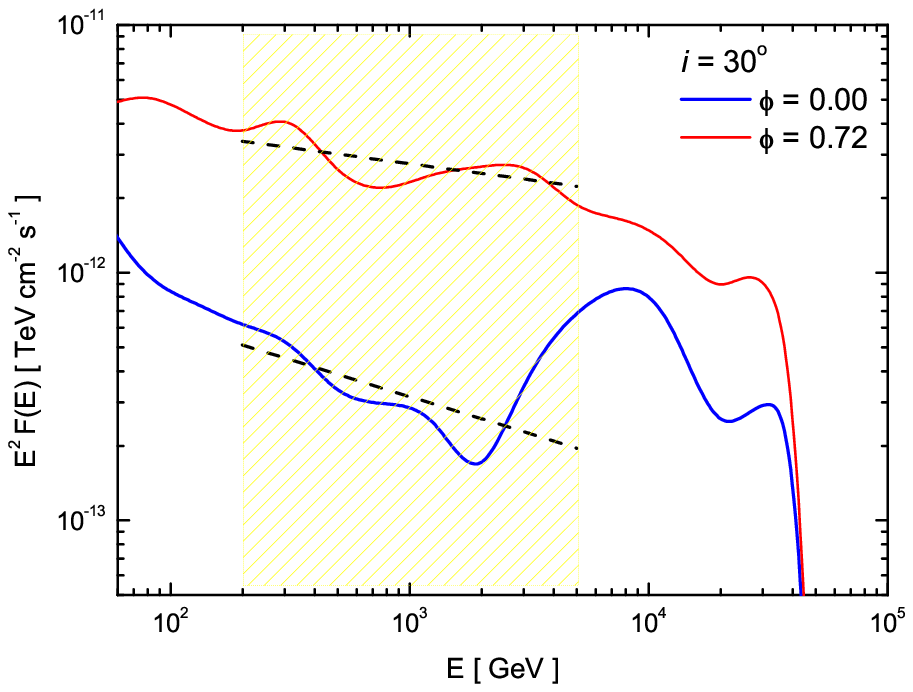}
 \includegraphics[width=.5\textwidth]{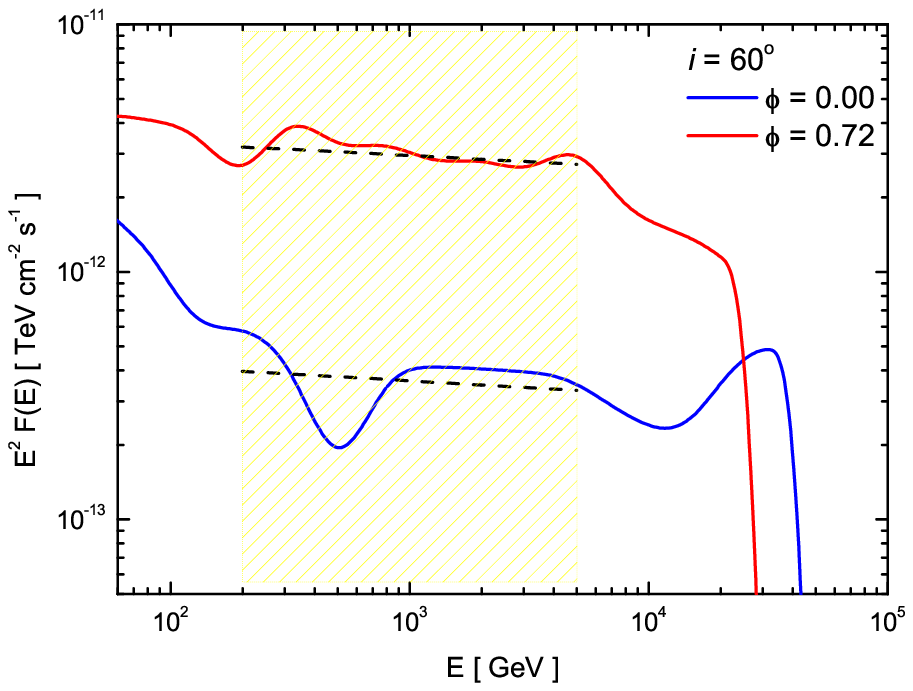}
\caption{Examples of the theoretical results, for different inclinations, for the spectrum of single phases fitted with a power law in the range 0.2--5 TeV. 
}
\label{single-spec}
\end{figure*}

Figure 1 shows the results of our model. In the top panels, we show lightcurves for photons with energy above 1 TeV for different binary inclinations (dot-dashed: $i=30^0$, solid lines: $i=60^0$), compared with  H.E.S.S. data. Black lines stand for results obtained with a variable interacting spectrum along the orbit; green (light) lines correspond to the constant spectrum case. In the middle panels, we show the pectra around INFC (in red) and SUPC (in blue). H.E.S.S. data is shown in the same colors. Results for both cases of interacting electron spectra are given. Shaded regions are energy ranges for which we study the lightcurves below. The two horizontal lines between 1 and 100 GeV represent sensitivities of GLAST in the all-sky survey mode. 
H.E.S.S. observations have also provided the evolution of the normalization and slope of a power-law fit to the 0.2--5 TeV data in 0.1 phase-binning (Aharonian et al. 2006). The variability of the spectral index and normalization of these fits along the orbit is impressive.  The use of a power-law fit and not a more complex higher-order curve has to do with low statistics in such shorter sub-orbital intervals: higher-order functional fittings (such as a power-law with exponential cutoff) provide a no better fit and were not justified.  We can directly compare with these H.E.S.S. results fitting our spectral predictions in the same way. 
In the bottom panels of Figure \ref{LC}, we show this fits with shaded areas:  in the left (right) panel,  we present the change in the normalization (photon index) of a power-law photon spectra fitted to the theoretical prediction for each of the 0.1 bins of phase. The two different colors of the shading stand for the two inclination angles considered. The size of the shading gives account of the error in the fitting parameters. Data points represent the H.E.S.S. results for the equal procedure: a power-law fit to the observational spectra obtained in the same phase binning.
We note that particularly for some phases close to periastron, the theoretical spectrum is badly mimicked by a power-law (see the examples of Figure \ref{single-spec}).  
Instead, for phases around INFC, power-law fits are a better description to our
model. 
Changes in the fitting parameters may also appear as a result of the  fitting range. 
e.g., fitting the theoretically computed spectrum with a power-law up to 5 or, 
say, 1 TeV introduce non-negligible changes. This is a non-trivial fact, due to possible scarcity of photons at high energies in reduced observation windows at different phases along the orbit, what should be confirmed with further observations.
The overall similarity of the theoretical and observational fittings is unexpected: we find that there is nothing a priori in the model that allows one to predict that when broad phase spectra data (INFC and SUPC) are reproduced so will be the data at the individual and much shorter phase-binning, less with such a good agreement. This result point perhaps to some 
reliability of the model, at least in its essential ingredients: geometry, cascading, interacting electron population.
Depending on inclination 
(see the power-law index for the case of $i=30^o$) 
there are a few points around INFC for which the observational data seem to be described by a harder spectrum than what we obtain as a result of fitting the theoretical predictions. The normalization data point close to periastron is missed by our predicted fitting ranges. This is the result of our spectrum being badly fitted by a power-law at these phases (phase 0 in Figure \ref{single-spec}).

\begin{figure*}
\centering
\includegraphics[width=.49\textwidth]{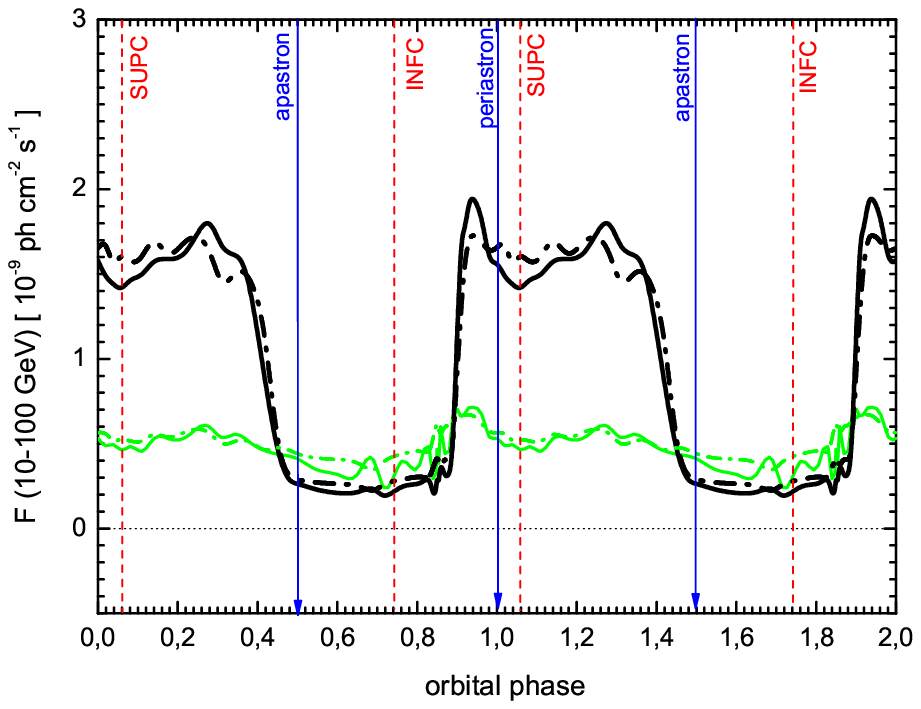}
\includegraphics[width=.49\textwidth]{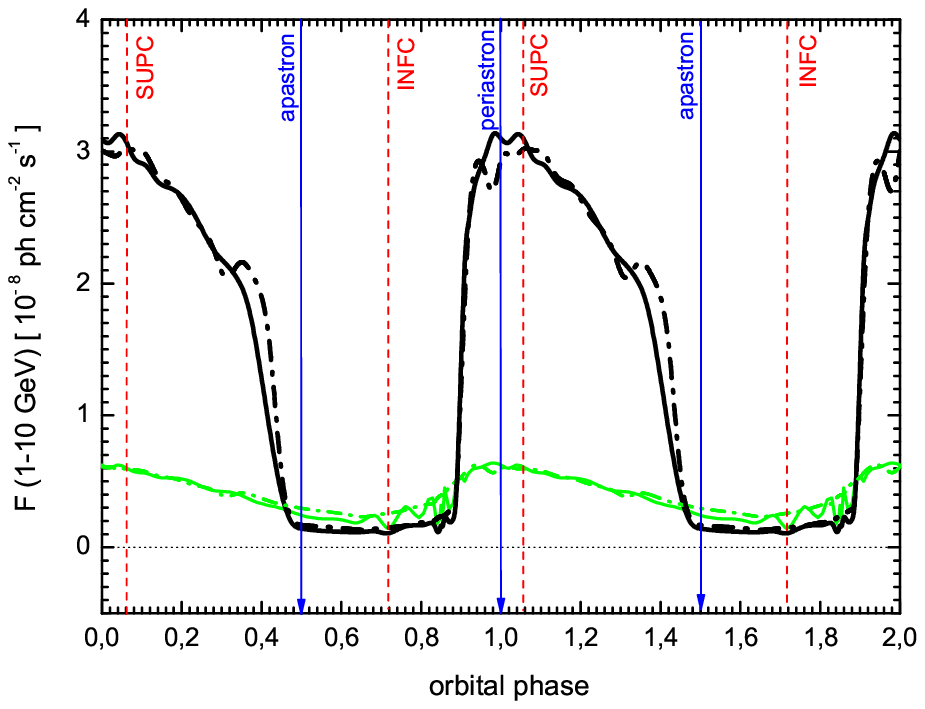}
\caption{Predicted theoretical lightcurves for the energy intervals 10 -- 100 GeV, and 1--10 GeV. Both inclination angles and interacting lepton spectra considered are shown  (dot-dashed: $i=30^0$, solid lines: $i=60^0$, black lines: variable spectrum of primary leptons, green (light) lines: constant spectrum along the orbit). 
}
\label{llcc}
\end{figure*}

\begin{figure*}[t]
\centering
\includegraphics[width=.49\textwidth]{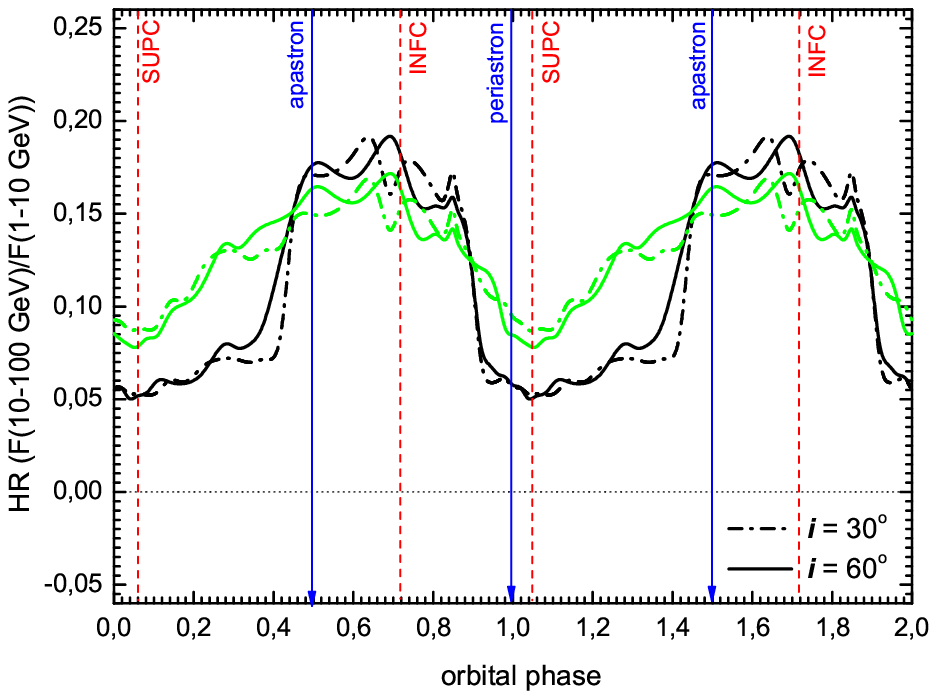}
\includegraphics[width=.49\textwidth]{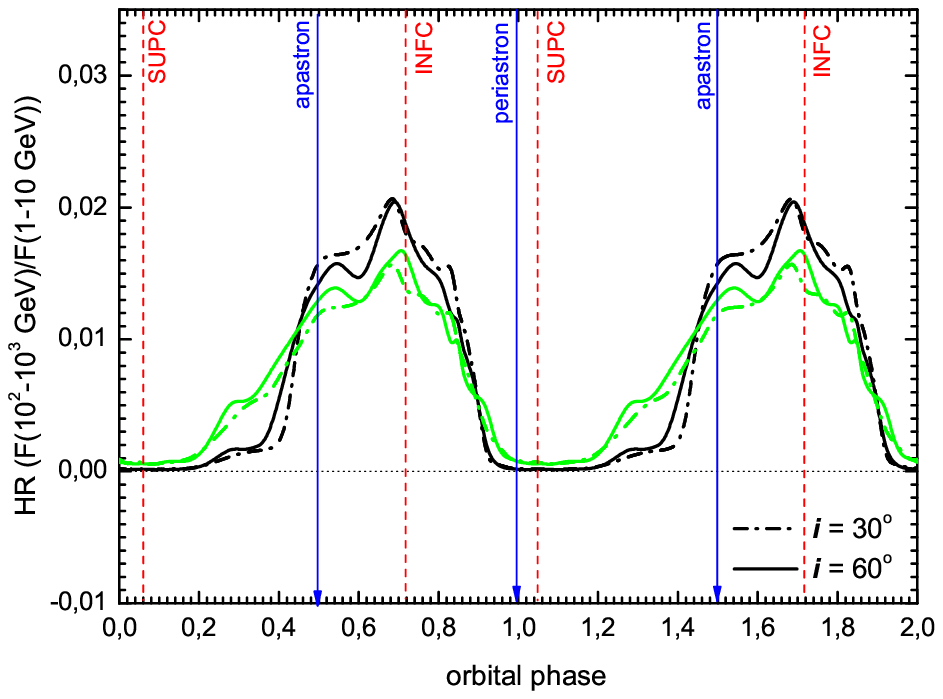}
\caption{Hardness ratios as a function of orbital phase, for two inclination angles (dot-dashed: $i=30^0$, solid lines: $i=60^0$). 
Color style follows that used in previous figures. The energy regimes  considered are $F({\rm 1 - 10 \, GeV})$, $F({\rm 10 - 100\, GeV})$, and $F({\rm 10^2 - 10^3 GeV})$. 
}
\label{HR}
\end{figure*}

Figure \ref{LC} also shows the spectral energy distribution predictions extended for energies above 1 GeV. The steeper the primary spectrum is, the higher the flux produced at lower energies, what is particularly notable for the SUPC broad phase-interval (see bold and thin lines of Fig.  1).
The (pre-launch) minimum flux
needed for a source to be detected by GLAST after a 1-month
and 1-year of operation in all-sky survey, for a $E^{-2}$ source.\footnote{www-glast.slac.stanford.edu/software/IS/glast\_lat\_performance.htm}
At these flux levels,  a 20\%-uncertainty in the determination of the flux, a resulting significance  about 8$\sigma$, and a spectral index determined to about 6\% would be achieved. Even when these sensitivities maybe slightly worse for a low-latitude source,  if this model is correct,  GLAST should be able to quantify the orbital variability after about a year of integration. The fact that the system periodicity is $\sim$3.9 days allows for a fast build-up of integration time around each of the portions of the orbit. A month-integration around SUPC will be obtained after $\sim$2 months of all-sky survey. LS 5039 is positionally coincident with 3EG J1824-1514, whose average $\gamma$-ray flux above 100 MeV --along all EGRET viewing periods-- 
is $\sim 3.5 \times 10^{-7}$ photons cm$^{-2}$ s$^{-1}$. Our model predictions 
for this energy range  are consistent with this observation too.
This model predicts a clear anti-correlation between the expected output at TeV and GeV energies, which can be seen comparing the lightcurves shown in Figure \ref{LC} and \ref{llcc}. The lower the energy range, the more anti-correlated the signal is with the TeV emission, what is a consequence of the phase-dependence of the cascading and absorption processes. A hardness ratio defined within the GLAST energy domain does not vary as much as it does when constructed combining the fluxes at low and high $\gamma$-ray energies (a factor of ~3 versus an order of magnitude), but it may be useful as a first check before integration time is achieved for determining the spectrum. This is shown in Figure \ref{HR}.

\begin{figure*}
\centering
\includegraphics[width=.49\textwidth]{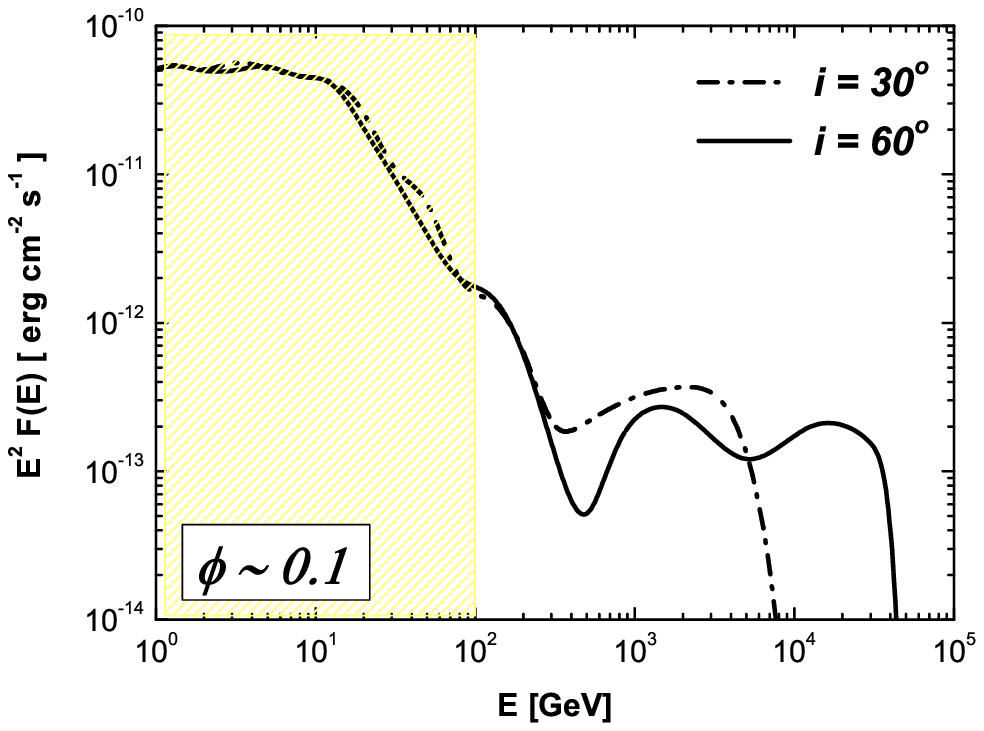}
\includegraphics[width=.49\textwidth]{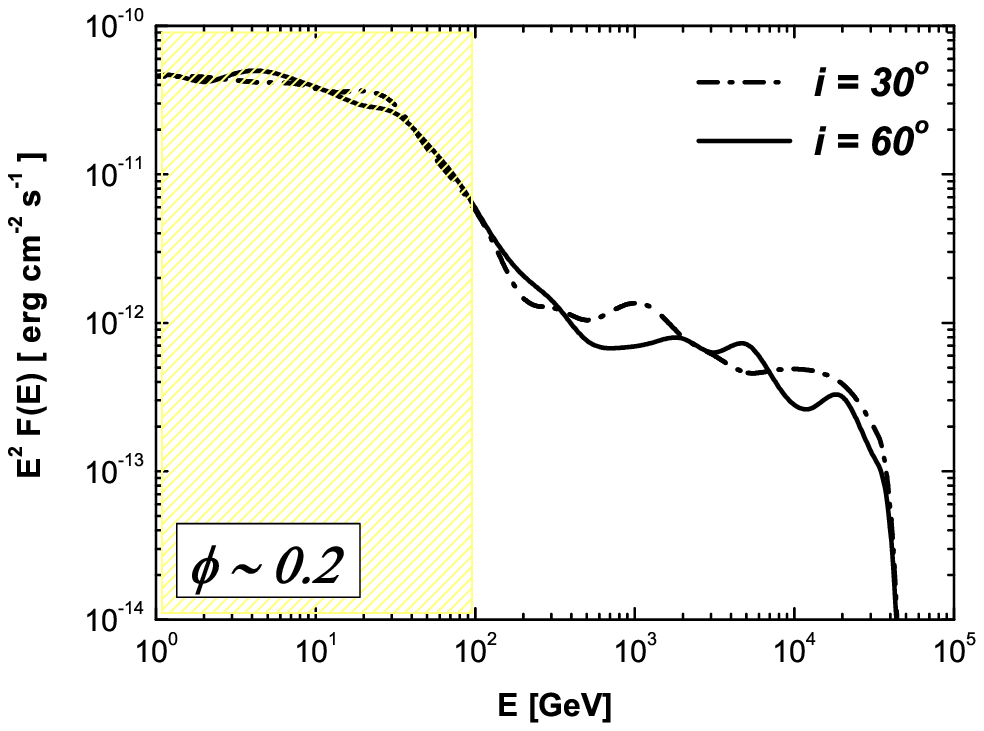}
\includegraphics[width=.49\textwidth]{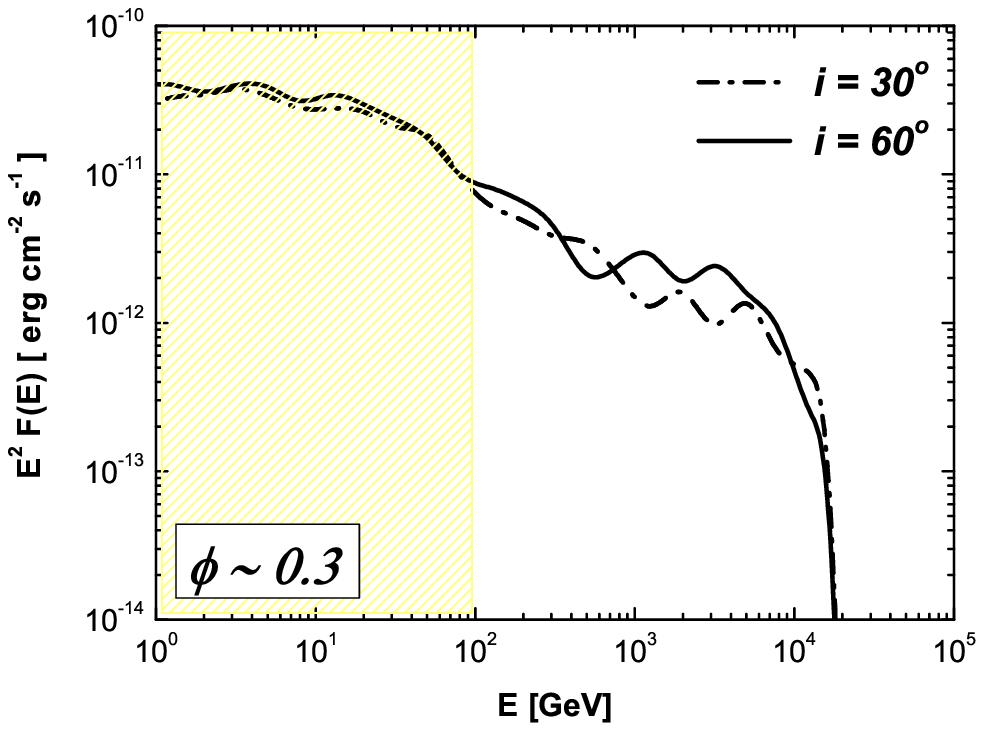}
\includegraphics[width=.49\textwidth]{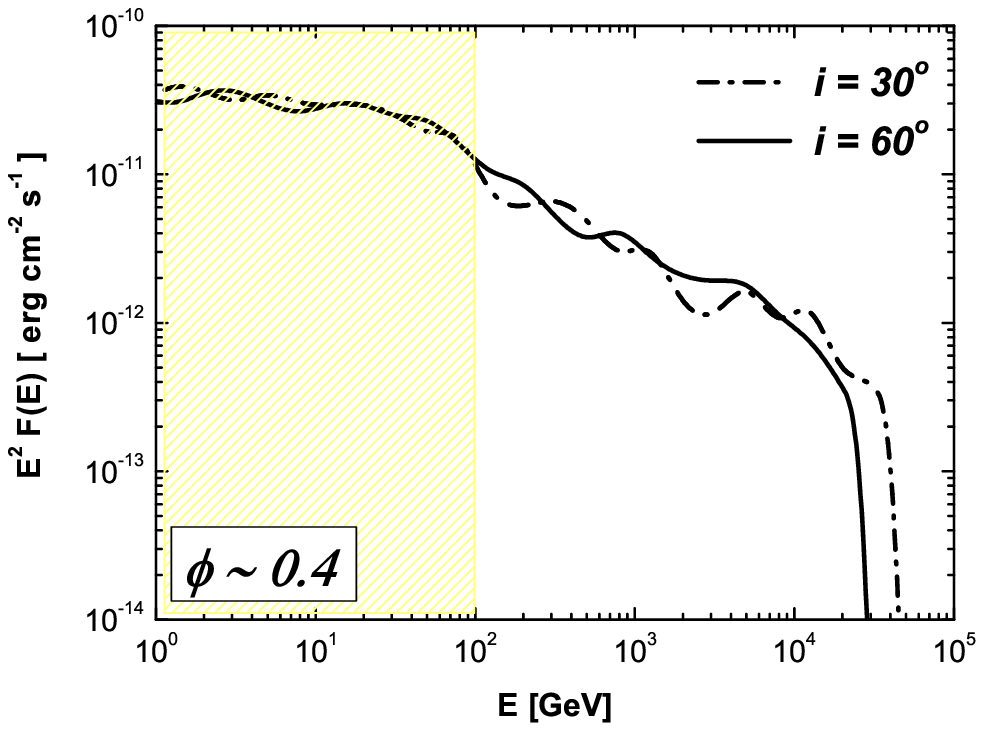}
\caption{Evolution of individual spectrum in the case of variable lepton distribution along the orbit, at individual phase bins, from 0.1 to 0.9. The shadow represents the GLAST energy coverage; the rest of the span of the x-axis can be observed by ground-based facilities.}
\label{phase-bin2}
\end{figure*}

\begin{figure*}
\centering
\includegraphics[width=.49\textwidth]{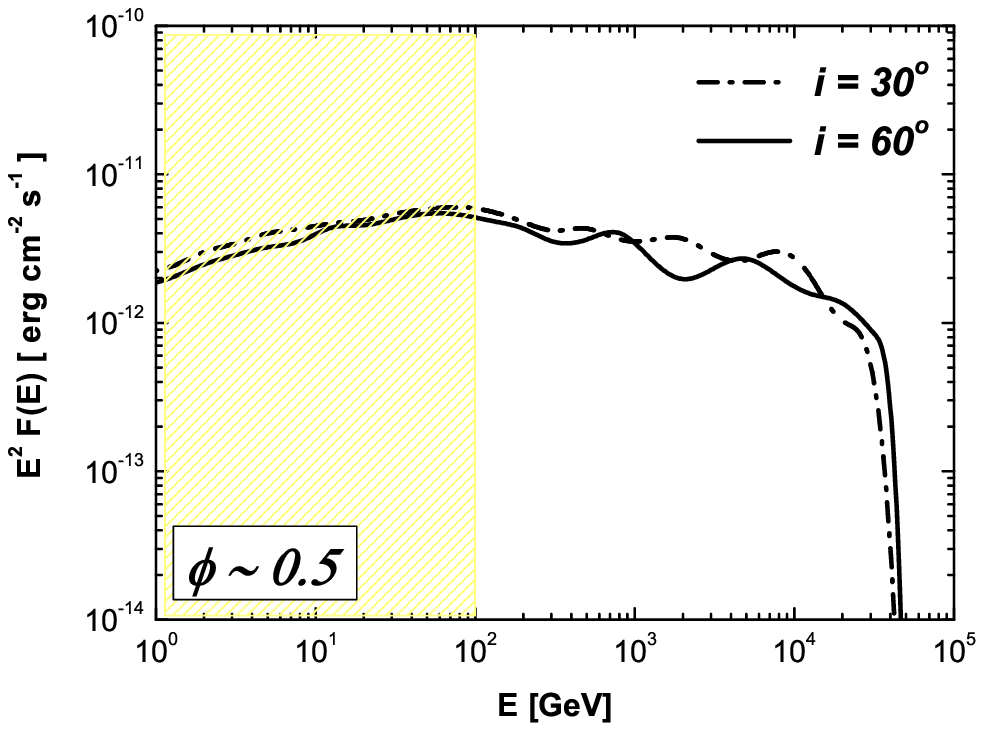}
\includegraphics[width=.49\textwidth]{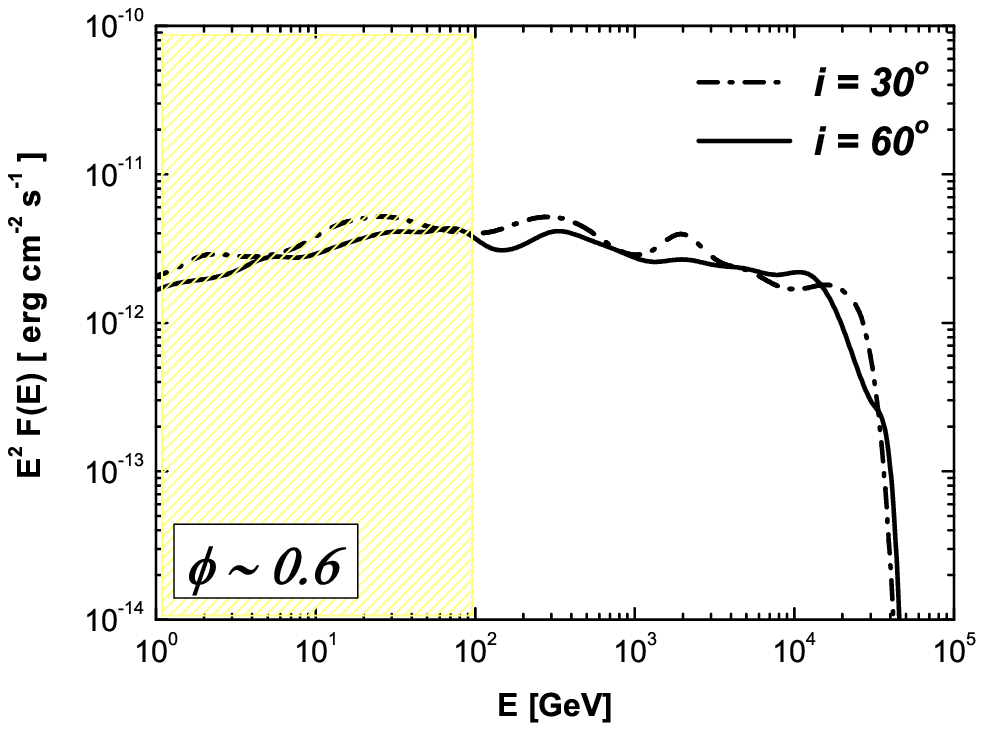}
\includegraphics[width=.49\textwidth]{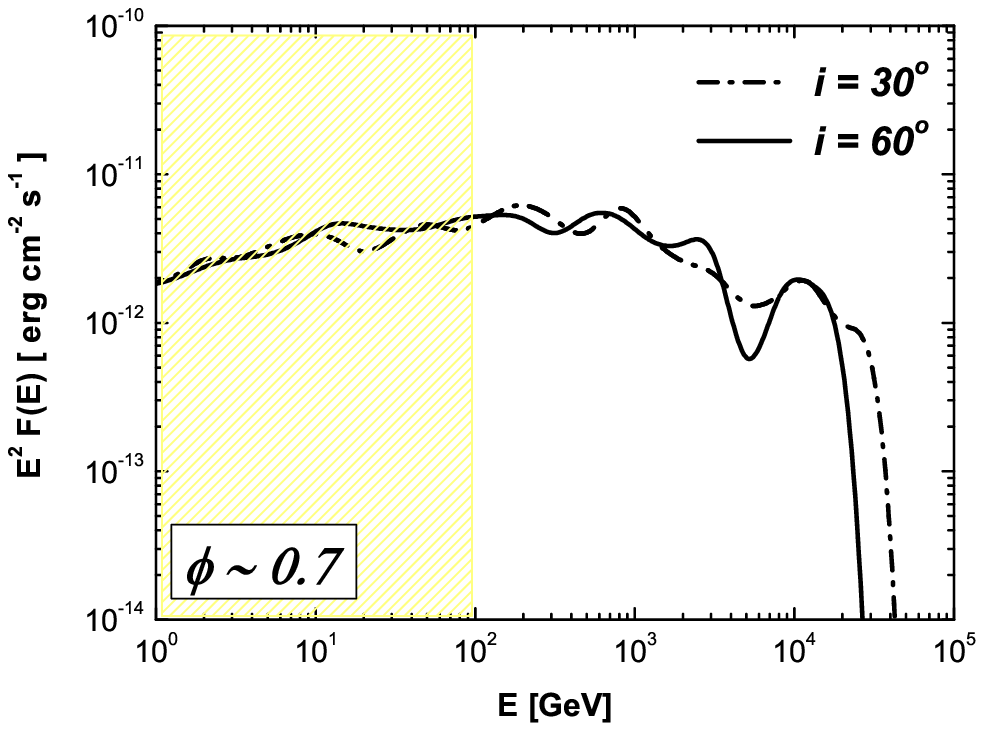}
\includegraphics[width=.49\textwidth]{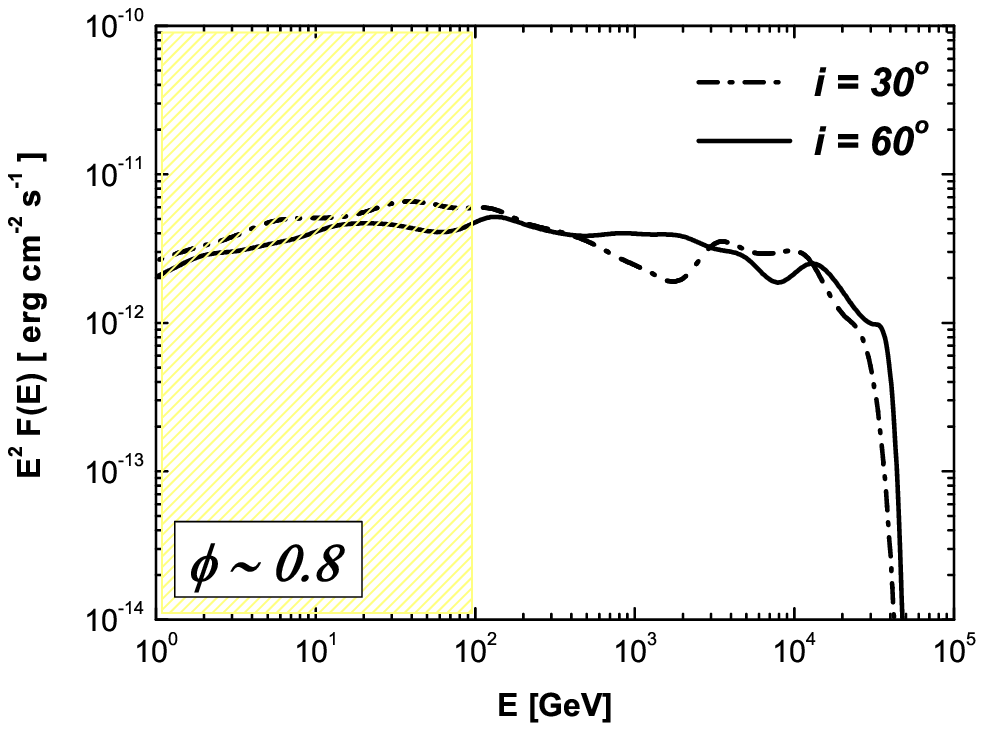}
\caption{Evolution of individual spectrum. Continued.}
\label{phase-bin22}
\end{figure*}

Apart from possible testing with GLAST (at the level of lightcurve, spectra, hardness ratios, and differentiation between constant and variable electron distribution), we can provide further possible tests at high $\gamma$-ray energies.
It was already said that power-laws do not always present the best fit to the  predicted spectra along the orbit.  Then, observations with larger statistics (with H.E.S.S., H.E.S.S. II, or CTA) could directly test the model in specific phases. A model failure in specific phases would allow further illumination about the physics of the system.
Figures \ref{phase-bin2}-\ref{phase-bin22} shows the evolution of individual spectrum in the best fitting case of variable lepton distribution along the orbit, at individual phase bins, from 0.1 to 0.9, and are useful for for testing with future quality of data. \\

This work was supported by grants AYA 2006-00530 and CSIC-PIE 200750I029.

\end{document}